\documentclass{aastex6}
\usepackage{nameref}
\usepackage{booktabs}
\usepackage{natbib}
\usepackage{amsmath,amstext}
\shorttitle{Latham 1 Abundances}
\shortauthors{O'Connell et al.}

\begin{document}

\title{Chemical Abundance Analysis of Moving Group W11450 (Latham 1)}

\author{
Julia E. O'Connell, 
Kylee Martens\altaffilmark{1,2}, and
Peter M. Frinchaboy
}
\affil{Department of Physics and Astronomy, Texas Christian University, TCU Box 298840, 
Fort Worth, Texas 76129, USA; j.oconnell@tcu.edu; p.frinchaboy@tcu.edu}


\altaffiltext{1}{Department of Physics, University of Wisconsin--Madison,
1150 University Avenue, Madison, WI 53706, USA;
kmmartens@wisc.edu}

\altaffiltext{2}{Department of Astronomy, University of Wisconsin--Madison,
475 N. Charter Street, Madison, WI 53706, USA}

\begin{abstract}

We present elemental abundances for all seven stars in Moving Group W11450 (Latham 1) to determine if they may be chemically related. These stars appear to be both spatially and kinematically related, but no spectroscopic abundance analysis exists in literature. Abundances for eight elements were derived via equivalent width analyses of high resolution (R$\sim$60,000), high signal--to--noise ratio ($\langle$SNR$\rangle\sim$100) spectra obtained with the Otto Struve 2.1m telescope and Sandiford Echelle Spectrograph at McDonald Observatory. The large star--to--star scatter in metallicity, -0.55 $\leq$ [Fe/H] $\leq$ 0.06 dex ($\sigma$= 0.25), implies these stars were not produced from the same chemically homogeneous molecular cloud, and are therefore not part of a remnant or open cluster as previously proposed. Prior to this analysis, it was suggested that two stars in the group, W11449 \& W11450, are possible wide binaries. The candidate wide binary pair show similar chemical abundance patterns with not only iron, but with other elements analyzed in this study, suggesting the proposed connection between these two stars may be real.\\

\end{abstract}


\section{INTRODUCTION}

Astronomers have long used star clusters as empirical testbeds for the purposes of understanding both the kinematic properties of the Galaxy as well as its chemical evolution. These bound stars share a common formation history, and as a result, share a common age, distance from us, relative speed, and spatial relationship on the sky. Open clusters are loosely bound, relatively young star clusters that reside almost exclusively in the plane of the Milky Way, making them particularly useful systems, in aggregate, for investigating Galactic dynamical and chemical evolutionary processes. Having been formed from the same well--mixed giant molecular cloud (GMC), they are also thought to be fairly homogeneous in their chemical composition  \citep[e.g][]{2006AJ....131..455D, 2007AJ....133.1161D}. For many elements, the observed star--to--star abundance variations for constituent members within open clusters show a remarkably small spread, $\sim$0.01 to 0.05 dex \citep[see][]{2010A&A...511A..56P, 2011MNRAS.415..563D, 2012MNRAS.427..882T, 2012MNRAS.419.1350R, 2013MNRAS.431.3338R, 2013MNRAS.431.1005D, 2016A&A...590A..74B}. These common attributes acquired from a shared formation history allow astronomers to differentiate stars that may belong to a particular cluster from those that happen to share the same field of view. It also allows for an examination of chemical evolutionary patterns that may emerge among these stars. \\

Even though the vast majority of stars, if not all, are born in clusters  \citep{2003ARA&A..41...57L}, the array of stars we see in the night sky are either single stars or binary systems, suggesting some disruption of clusters must have occurred on a relatively short timescale \citep{2003ARA&A..41...57L,2005A&A...443...79B,2005A&A...443...41M}. The Galactic disk is a dynamic environment where close encounters or interactions with other clusters, stars in the field, or GMCs can disrupt a cluster and cause it to lose members to the disk. Perturbations due to spiral arm rotation and the central Galactic bar may cause orbital variations that can disperse cluster members, and orbital resonances that may trap dispersed member stars \citep{1998AJ....115.2384D}. Both internal star formation dynamics \citep[see][for a review]{2010RSPTA.368..713L} and external Galactic perturbations can oust cluster member stars where they migrate to the disk as unbound stars, or moving groups, and become part of the general stellar population. While Galactic orbital parameters may change for these stars their chemical composition is preserved, as well as the small star--to--star elemental scatter \citep[see][]{2007AJ....133..694D, 2010AJ....140..293B}.  High resolution, high SNR spectroscopic abundance analyses can reveal the chemical signatures necessary for identifying such stellar associations for the purpose of ``chemically tagging"  them to either the thick or thin disk of the Galaxy \citep{2002ARA&A..40..487F}.  While beyond the scope of this study, chemically tagging stars to their Galactic zip code, in the interest of reconstructing the chemical evolution of the Galaxy, is an intriguing proposition for near field cosmology.\\

In a systematic radial velocity survey centered on ``Selected Area 57"\footnote{The term is an historical holdover from a plan originally proposed by Jacobus Cornelius Kapteyn in 1906 to determine the structure of the Galaxy by taking samples at regularly spaced intervals.\citep[see][]{1906BuAsI..23..480.}.}(SA 57; \emph{l}=66$\degr$, \emph{b}=+86$\degr$)  near the North Galactic Pole, \citet[hereafter LMS91]{1991AJ....101..625L} noted seven specific stars, referred to as Group W11450, located within a circle of radius $0.5 ^{\circ}$ near the edge of the survey region with a small velocity dispersion of $\sim$0.27 km s$\rm ^{-1}$. The group gets its name as a consequence of a high--probability bound pair residing within the stellar association, W11450AB--W11449, the `AB' indicating a possible tertiary system \citep{1988Ap&SS.142..131M}.  As described by \citet{1984ApJ...281L..41L}, distances for four LMS91 stars are estimated by calibrating absolute magnitudes \emph{v.} spectral types, which the study also provides. LMS91 affirms the distances found are consistent with all stars being within 2 pc of each other, commensurate with an open cluster tidal radius of $\sim$10 pc \citep{1998A&A...331...81P, 1998A&A...329..101R, 1999A&A...345..471R}.  By extrapolating masses down to late M stars, and applying the angular diameter of the ``cluster" together with the velocity dispersion, the study concludes that the moving group is virialized and therefore bound, or close to being bound. \\

Listed as Latham 1, and hereafter referred to as such, in the current Dias catalog of open clusters \citep{2002A&A...389..871D}\footnote{The catalog can be found online at http://www.astro.iag.usp.br/ocdb}, a search of current literature reveals few details about these stars. Interestingly, a study of blue horizontal branch (BHB) halo stars in SA 57 from \citet{1994AJ....108.1722K} considers a Latham 1 star, W13284 identified as Case A-F 860, in the initial star sample. All seven Latham 1 stars have reliable 2MASS \citep[The Two Micron All Sky Survey;][]{2006AJ....131.1163S} \emph{JHK} photometry and $B_TV_T$ colors from the Tycho-2 catalog \citep{2000A&A...355L..27H}. By leveraging both 2MASS and Tycho surveys, \citet{2006ApJ...638.1004A} produced photometric estimates of stellar parameters, including metallicities for four stars and effective temperatures for all stars considered in this study. Utilizing the same surveys, \citet{Casagrande} contributes a metallicity estimate for one star, W23375, which is common to both studies. This star also appears in the \emph{Hipparcos} catalog \citep{1997A&A...323L..49P} with a parallax distance consistent with LMS91, 108 pc and 101 pc, respectively. While both \citet{2006ApJ...638.1004A} and \citet{2010PASP..122.1437P} provide distances for all Latham 1 stars, the two studies show considerable differences in their estimates. \citet{2010PASP..122.1437P} also provide spectral and luminosity classes, revealing all seven stars are dwarfs ranging in temperature from F0 to G8. We summarize the estimates of distances, metallicities and spectral types from these earlier studies in Table \ref{tab1}.  \\

Withal, Latham 1 has a number of essential parameters that remain unmeasured. In an effort to remedy this deficiency we conducted a study of Latham 1 stars to derive their metallicities. Therefore, we present the first high resolution spectroscopic study for abundances of Fe I \& Fe II (neutral and singly ionized iron, respectively), Na I, Mg I, Si I, Ca I, Ni I and Ba II in an effort to determine whether Latham 1 stars have individual chemical compositions consistent with the group being defined as a cluster, or remnant cluster.\\

In \S{2} we describe the observations and data reduction sequence for program stars in this study. Radial velocity measurements are presented in \S{3}. In \S{4} we take up the procedure for estimating model stellar atmospheres and for measuring the chemical abundances. We state results and provide some discussion in \S{5}, and conclude with a summary of this study in \S{6}.\\\\

\section{OBSERVATIONS AND DATA REDUCTION}

Observations for all target stars were obtained at McDonald Observatory with the Otto Struve 2.1m telescope instrumented with the Sandiford Cass Echelle Spectrometer (SES; R=$\lambda$/$\Delta$$\lambda$$\approx$60,000) in June 2013 and 2014. The spectrograph setup was centered near 6175~\AA\ in 2013 and  6375~\AA\ in 2014 with wavelength coverage spanning from $\sim$5350--6560~\AA\ and $\sim$5560--6750~\AA, respectively (see Table \ref{tab2} for observation dates and exposure times). Colors for all stars were taken from the Tycho-- 2 catalog \citep{2000A&A...355L..27H} and corrected to Johnson photometric magnitudes by the standard $V   = V_T - 0.090\times(B_T-V_T)$ and $B-V = 0.850\times(B_T-V_T)$. Program stars are located near the North Galactic Cap in SA 57 (\emph{l}=66$\degr$, \emph{b}=+86$\degr$) and range in magnitude from $8.25 \leq m_V\leq 11.47$ after corrections. Updated coordinates were provided by the SIMBAD astronomical database, as LMS91 provides coordinates from 1950. Both LMS91 and \citet{1981ApJ...246..122B} reference the Weistrop (North Galactic Pole) 1980 catalog by private communication for star identification, including  W11449 \& W11450. Identification for both stars based on Weistrop numbers from SIMBAD disagree with their coordinates. However,  LMS91 correctly identifies W11449 as ADS 8811 C, and W11450 as ADS 8811 AB from \citet{1932ngcd.book.....A}, so we have based our updated coordinates on those positions instead.\\ 

Basic data reductions were carried out using the standard IRAF\footnote{IRAF is distributed by the National Optical Astronomy Observatories, which are operated by the Association of Universities for Research in Astronomy, Inc., under cooperative agreement with the National Science Foundation.} routines. Specifically, \emph{ccdproc} was used to apply the bias level correction, trim the overscan region and apply a normalized flat to object data and comparison arcs.  Due to the long exposure times, L.A. Cosmic \citep{2001PASP..113.1420V} was introduced into the reduction sequence for the detection and removal of cosmic rays without altering the flux level of the bias--subtracted, un--scattered light subtracted data. The IRAF package \emph{echelle} was used for aperture tracing, scattered light removal, extraction of the one--dimensional spectra,  flat--fielding and wavelength calibration (based on a Th--Ar comparison source). The extracted spectra were continuum fit using a low order spline function then co--added to increase the signal--to--noise ratios (SNRs) of the final spectra. The SNRs of the combined spectra range from $\sim$70--130. \\

\section{RADIAL VELOCITIES}

All radial velocities for this study were determined by using the Fourier cross--correlation function provided by the IRAF task \emph{fxcor}, and corrected for the Earth's rotation using \emph{rvcorrect}.  Four Latham 1 stars (W12388, W13284, W21415 and W23508) are included in a radial velocity study from \citet{2006PASP..118.1656S}, who recommends these stars as secondary velocity standards. Table \ref{tab3} presents our radial velocity measurements and associated uncertainties for each program star, as well as a comparison to radial velocities found by LMS91 and \citet{2006PASP..118.1656S}. Our results reproduced the small velocity dispersion identified by LSM91, $\sim$0.33 km $\rm s^{-1}$, which is on the order of the typical measured open cluster velocity dispersion, $\sim$1 km $\rm s^{-1}$ \citep[e.g.][]{2010AJ....139.1383G,2015AJ....150...97G}. \\

For W11450, we note that LMS91 estimates the center of mass velocity for a possible tertiary system described by \cite{1988Ap&SS.142..131M}. As for W23375, the hottest star in the group, absorption lines of singly ionized atoms dominate the spectrum and fewer ``clean"  lines of neutral species are available for measuring both radial velocity and elemental abundances. For this reason we assign the greatest uncertainty in our velocity measurement. Moreover, we chose to measure only the iron content for this star.\\

\section{ABUNDANCE ANALYSIS}

We have analyzed Latham 1 stars for elemental abundances of  Fe I \& Fe II (neutral and singly ionized iron, respectively), Na I, Mg I, Si I, Ca I, Ni I and Ba II, with the exception of W23375. For this star we have derived iron abundances only. All elemental abundances were derived using an equivalent width (EW) analysis. The IRAF task \emph{splot} was used to measure EWs with a single line, gaussian fit for unblended lines or, when necessary, a blended-line function for heavily blended lines or lines subject to hyperfine splitting. The wavelength range of observed spectra covers $\sim$5350--6560~\AA\ for stars observed in 2013, and $\sim$5560--6750~\AA\ for stars observed in 2014. Effective temperatures (T$_{\rm eff}$) for individual stars were initially estimated by utilizing photometry provided by the Tycho-- 2 catalog, which provides  $V_T$ and $B_T$ colors for all stars from a single source. For initial estimates of surface gravities (log $g$), we adopted distances for individual stars from \citet{2010PASP..122.1437P}, which made use of the proper motions listed in the Tycho--2 catalog to calculate distance estimates (see their \S 5.2 for details). The LTE stellar line analysis code MOOG \citep{1973PhDT.......180S} was then used to finalize model atmosphere parameters, T$_{\rm eff}$, log $g$, and microturbulence ($\xi$), and to derive elemental abundances.\\

\subsection{Model Stellar Atmospheres}

\begin{figure}
\caption{Filled circles indicate photometric estimates of effective temperature and our final T$_{\rm eff}$ derived spectroscopically. The dashed diagonal represents perfect agreement.}
\hspace{0.05in}
\epsscale{0.95}
\plotone{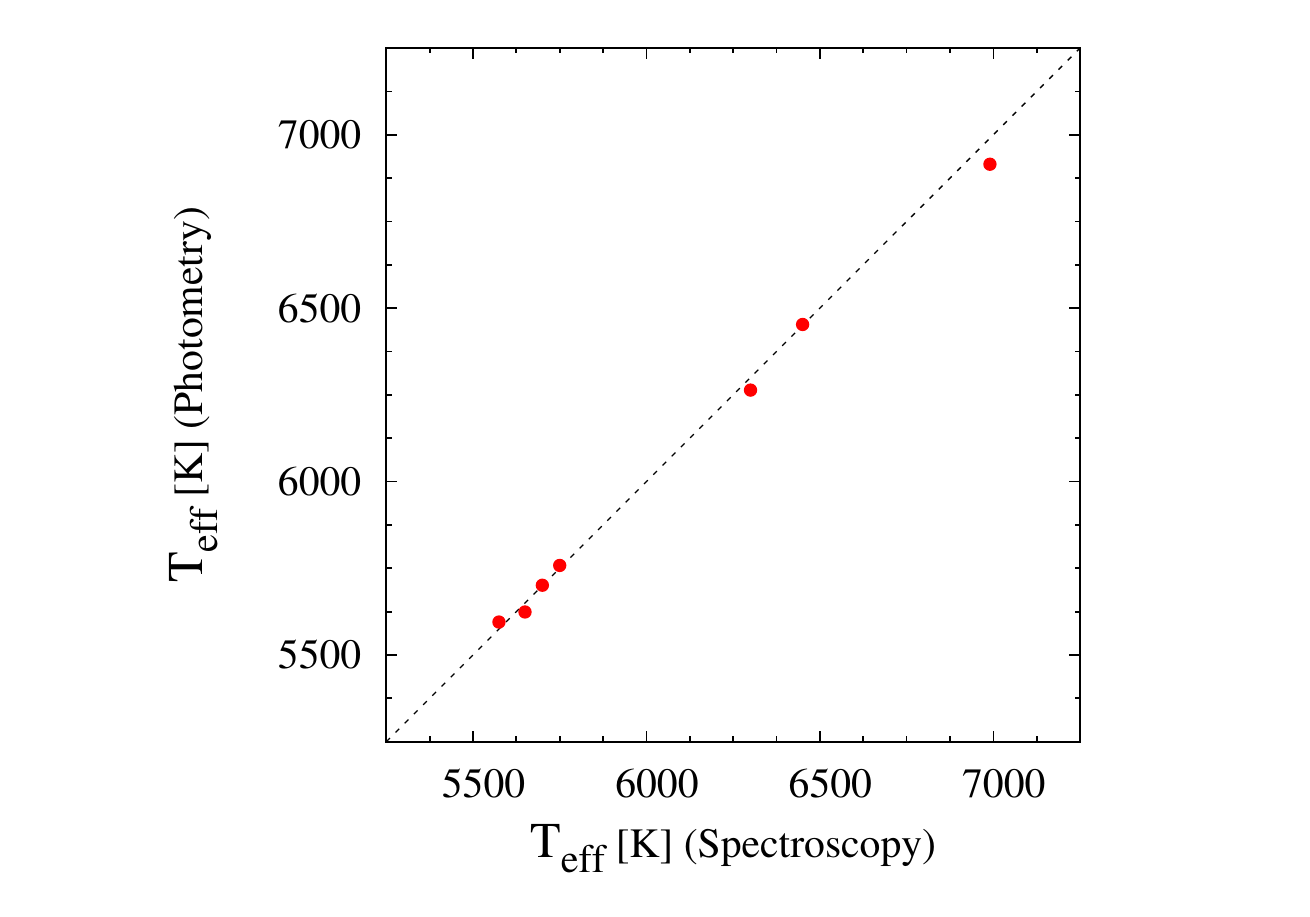}
\label{f1}
\end{figure}

Initial estimates of T$_{\rm eff}$ were determined for individual stars by utilizing the corrected $(B_T-V_T)$ color indices and color--temperature relationship described by \citet{2000AJ....120.1072S} for the their full sample. The Tycho-- 2 magnitudes were converted to the Johnson photometric system and corrected for interstellar reddening to provide the appropriate $(B-V)_0$ color indices. The dust maps provided by both \citet{2011ApJ...737..103S} and \citet{1998ApJ...500..525S}, accessed through the NED Coordinate Transformation \& Galactic Extinction Calculator,\footnote{https://ned.ipac.caltech.edu/forms/calculator.html} reveal a color excess E(B--V) value no greater than 0.01 for each star. The correction for interstellar reddening and extinction, though almost negligible, was performed to render the appropriate $(B-V)_0$ prescribed by Sekiguchi and Fukugita, specifically their equation 2, which includes a weak dependence on [Fe/H]\footnote{We adopt the standard notations [A/B]$\equiv$log(N$_{\rm A}$/N$_{\rm B}$)$_{\rm star}$--log(N$_{\rm A}$/N$_{\rm B}$)$_{\sun}$ and log $\epsilon$(A)$\equiv$log(N$_{\rm
A}$/N$_{\rm H}$)+12.0 for elements A and B.}. As literature provides few estimates of [Fe/H] for these stars (see Table \ref{tab1} for details),  and none for all stars from a single source, we initially assumed solar metallicity, where log$\epsilon$(Fe) = 7.50 \citep{2009ARA&A..47..481A}. This provided for only a rough estimate of initial model temperatures. With [Fe/H] derived we find a near 1:1 correlation between photometric estimates and spectroscopic T$_{\rm eff}$ for all program stars (see Figure \ref{f1}). \\

\begin{figure}
\caption{Plotting designations are the same as Fig. \ref{f1}, but for photometric estimates of surface gravities and log $g$ determined by ionization equilibrium. The inverted triangle represents an estimate of log $g$ for W13284 calculated at a distance of 1000 pc and our final surface gravity determined for this star; the dot-dashed line connects the two estimates.}
\hspace{-0.10in}
\epsscale{0.93}
\plotone{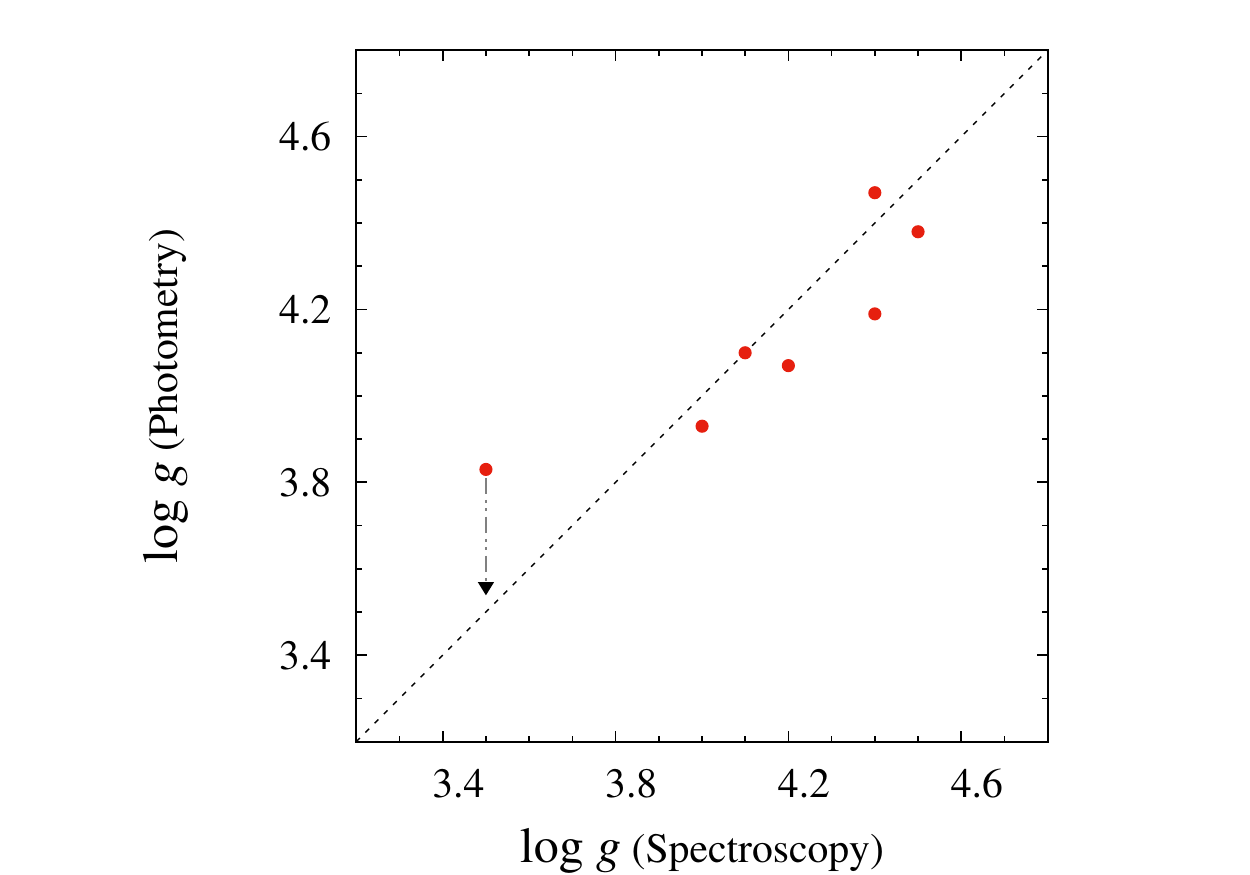}
\label{f2} 
\end{figure}

Surface gravities were initially estimated from the standard relation, 
\begin{equation}
 log(g) = 0.40(M_{bol.}-M_{bol.\sun}+log(g_\sun)+4(log(T/T_\sun))+log(M/M_\sun),
\end{equation} and assumed a stellar mass of of 1 M$_{\sun}$. The absolute bolometric magnitudes (M$_{\rm bol.}$) were determined by applying the $V$--magnitude bolometric correction from \citet[Appendix D]{1998A&A...333..231B} to the absolute $V$--band magnitudes estimated from distance moduli ($M-m$) for individual stars.  Surface gravities were further refined by an iterative process that focused on balancing Fe II with Fe I abundances.  Figure \ref{f2} shows the correlation between our initial estimates of surface gravity and our final log $g$ derived spectroscopically. As previously mentioned, W13284 was considered in the initial sample of stars for a study of BHB halo stars in SA 57. Although not included in their final sample, we investigated a log $g$ estimate calculated at a distance of 1000 pc for this star. As a note of interest, we find a good correlation for our spectroscopically derived log $g$ and the photometric log $g$ estimate determined at this distance, which we include in Fig. \ref{f2}. \\

All stellar atmospheres were modeled by interpolating in the ATLAS9 grid\footnote{Kurucz model atmospheres can be found at http://kurucz.harvard.edu/grids.html} and new opacity distribution functions \citep[ODFNEW,][]{2004astro.ph..5087C} for use with the line analysis code MOOG. Final T$_{\rm eff}$ were modeled by removing trends in Fe I with excitation potential. Surface gravities were further constrained by Fe ionization equilibrium. We initially assumed a $\xi$ value of 2 km $\rm s^{-1}$ for all stars, and achieved a final value by eliminating any remaining bias in Fe I with reduced EW [log(EW/$\lambda$)]. Elemental abundances were determined when a final solution over all parameters was achieved. Our final model atmospheres, photometry and SNR of the co--added spectra can be found in Table \ref{tab4}.\\

\subsection{Equivalent Width Analysis and Atomic Data}

All element abundances were derived by EW measurements using IRAF\rq{}s \emph{splot} package and by utilizing the \emph{abfind} driver in the 2014 version of the line analysis code MOOG. We make the assumption that LTE conditions predominate. 
Lines were chosen for measurement both by visual inspection and by comparison to the \citet{2000vnia.book.....H} Arcturus atlas, which provides  line profiles for a high--resolution solar spectrum in addition to the Arcturus spectrum. Given the high resolution of our spectra, we chose lines for analysis that were not expected to be severely blended. However, consistently strong, generally $>$ 170 m\AA, or obviously blended lines were discarded during analysis. This was especially true for star W23375, which may be a fast rotator. As a moderately warm, $\sim$7000 K, F-- type star with the lowest SNR of the group, ionized lines tend to blend with nearby neutral lines, or suffer from line blanketing effects altogether, making few lines available for reliable measurement. Therefore, we have restricted our abundance analysis for this star to Fe I (eight lines) and Fe II (two lines). Even so, the results for this star should be regarded with some skepticism; we include it for purposes of completeness.  \\

The SES suffers from internal reflection, referred to as the ``picket fence", which affects $\sim$20 \AA\ in one order per raw image, the order being dependent on the grating angle setting, which in turn determines the wavelength range. Because the spectrograph rides on the back of the telescope, flexure from its weight at zenith distances $>$$\sim$ 30 degrees can shift the exact location of the picket fence. For stars observed in 2013, i.e. W21415, W23375 \& W23508,  the picket fence affects $\lambda\lambda\sim$ 5975--5995 \AA, which renders the Fe  I $\lambda$ 5987 \AA\ line unmeasurable. For stars observed  in 2014, i.e. W11449, W11450, W12388 \& W13284, the picket fence appears at $\lambda\lambda\sim$ 6315--6335 \AA. Generally affected lines are Fe I $\lambda$ 6322 \AA\ and all Mg I lines in the $\lambda$ 6300 \AA\ region. Even so, the picket fence  shifted sufficiently during observations of W12388 \& W13284 to allow measurements of Mg I $\lambda$ 6318 \AA\ \& Fe I $\lambda$ 6336 \AA, respectively.\\

Analyses of stellar abundances are dependent on quality atomic data suitable for the stars in question. The line list was created with the goal of having the best abundance determinations given the array of  T$_{\rm eff}$  estimates for stars in this study, and the wavelength ranges provided from our observations. Where self--consistent we have used atomic data from recent studies, otherwise adopted older, yet still reliable, oscillator strengths. 
Atomic data for Fe I were culled from recent lab studies by \citet{2014MNRAS.441.3127R} and \citet{2014ApJS..215...23D}, with the exception of  $\lambda\lambda$ 5775 \& 6056 \AA\ from \citet{1991JOSAB...8.1185O}. For Fe II we adopt $gf$--values from Kurucz\footnote{See http://kurucz.harvard.edu/linelists.html, specifically the atomic data in gfnew (and gfall for the ``older" data),\\ and references contained therein.} rather than the more general oscillator strengths from the NIST database. The recently available atomic data from Kurucz was also used for Na I, Mg I and Ca I, and for Si I with excitation transitions $>\sim$5.6 eV. For Si I with excitation potentials $\sim$5.1 eV or lower, the ``older" log $gf$'s from the Kurucz database returned more self--consistent results. For Ni I, log $gf$ values  for $\lambda\lambda$ 6176, 6204 and 6223 \AA\ were taken from \citet{2014ApJS..211...20W};  \citet{1988JPCRD..17S....F} provided values for $\lambda\lambda$ 5593 \& 5760 \AA; and for $\lambda$ 5805 \AA\ \citet{Kurucz} was adopted. Since Ba II is subject to hyperfine structure and/or isotopic broadening, we utilized the log $gf$'s following \citet{1998AJ....115.1640M}. The final line list with EW measurements can be found in Table \ref{tab5}.    \\

\section{RESULTS AND DISCUSSION}

\begin{figure}
\caption{A comparison of Fe I \& II measured in all seven stars. Error bars indicate the line--to--line deviation in the measured abundance. The dashed line represents solar metallicity. }
\epsscale{1.00}
\hspace{0.05 in}
\plotone{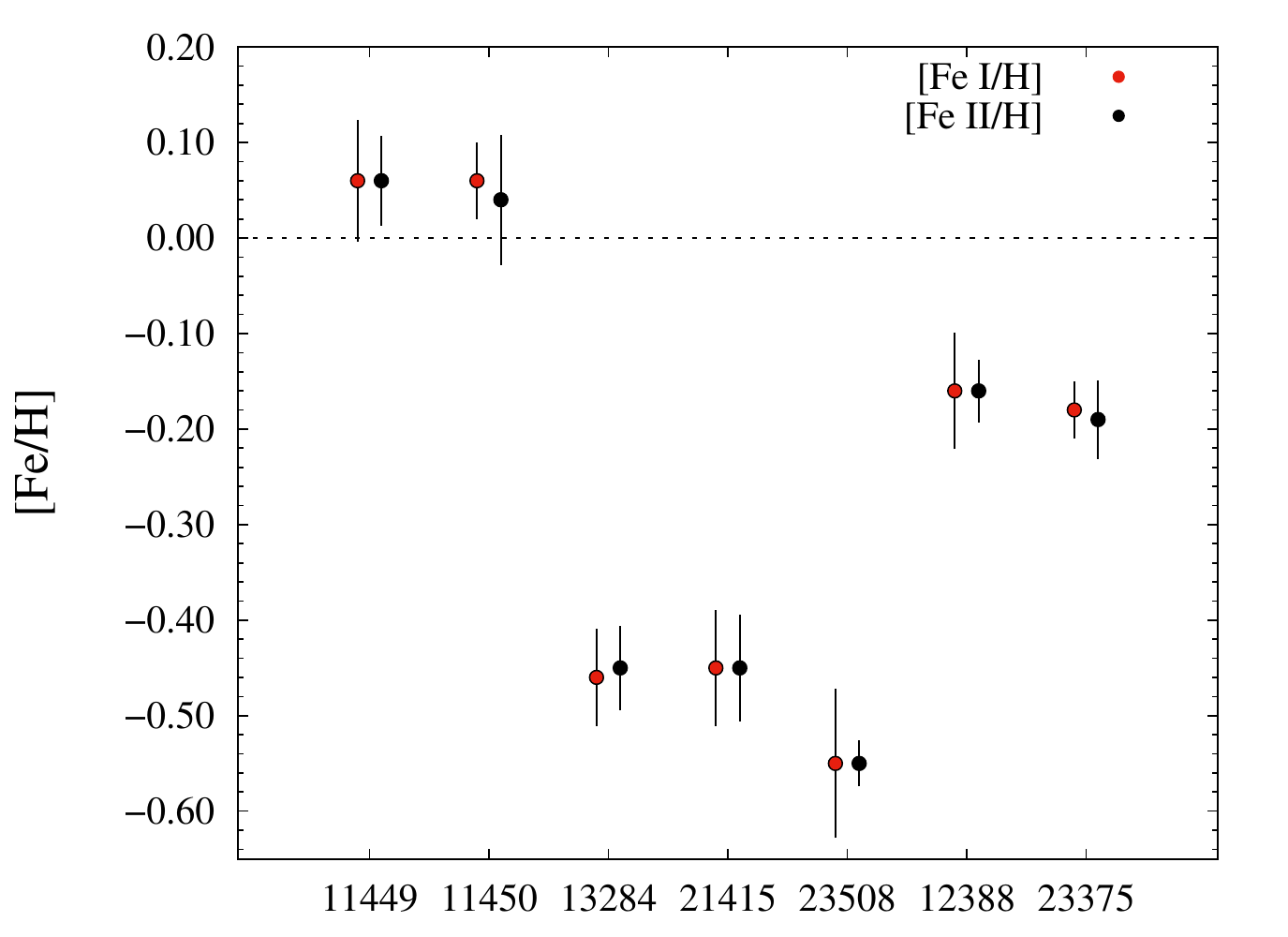}
\label{f3}
\end{figure}

\begin{figure}
\caption{Elemental abundances for stars in the [Fe/H] $\sim$ -0.50 dex range. Plotting designations are the same as Figure \ref{f3}. While the stars share similar [Fe/H], the large star--to--star dispersion in abundances suggests these stars are chemically unrelated. }
\epsscale{1.00}
\plotone{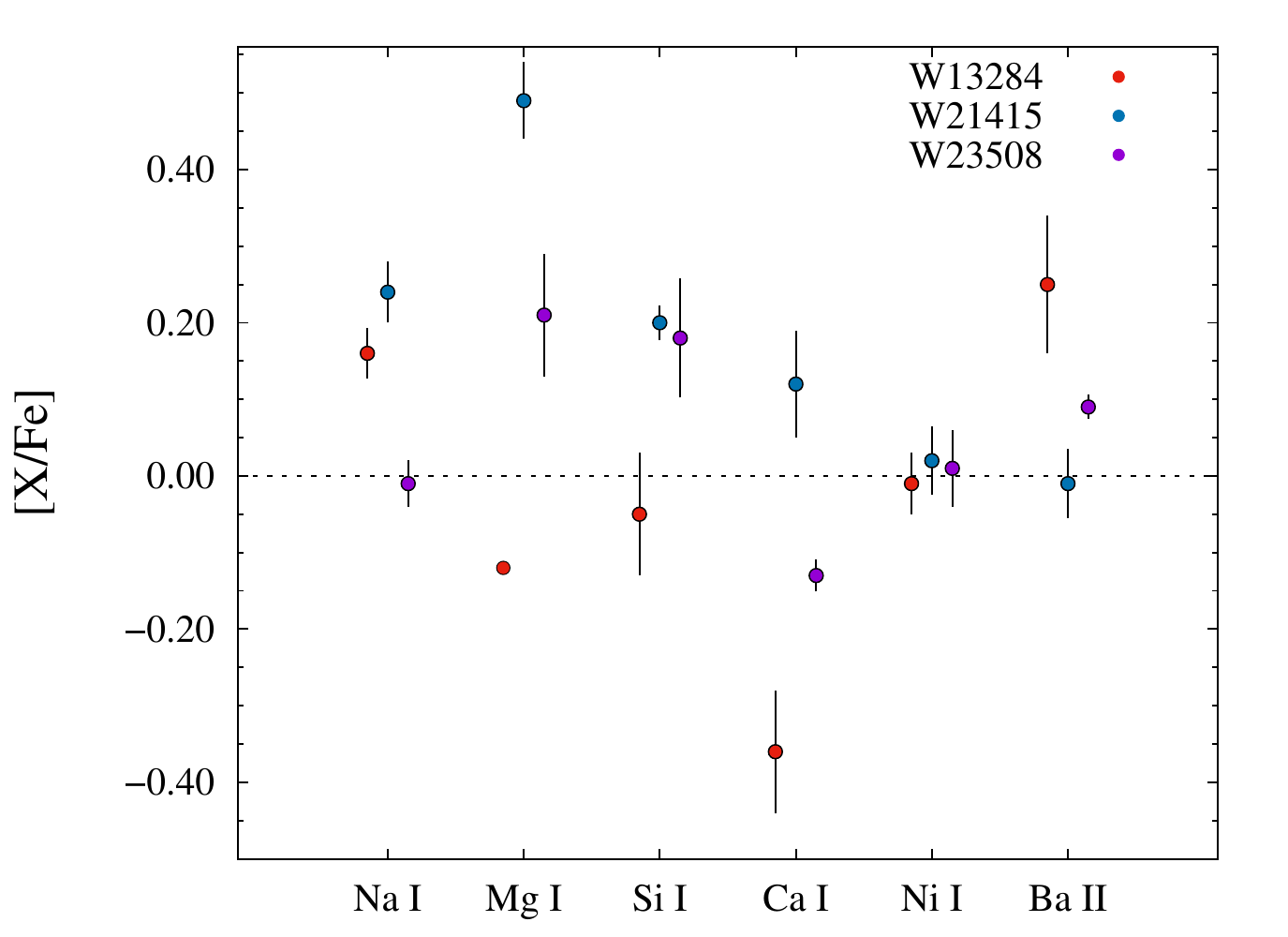}
\label{f4}
\end{figure}

We have derived chemical abundances for seven stars that constitute Moving Group W11450, or Latham 1, in an effort to determine whether they may be chemically related. All measured abundances can be found in Table \ref{tab6}. The large star--to--star scatter in metallicity, -0.55 $\leq$ [Fe/H] $\leq$ 0.06 ($\sigma$= 0.25),  precludes all stars being formed from the same chemically homogeneous GMC. Ad hoc examination of the $>\,\sim$0.60 dex  spread in [Fe/H] values displayed in Figure \ref{f3} appears to segregate the stars into three distinct groups:  [Fe/H]$\sim$ -0.17, [Fe/H]$\sim$ -0.50, and [Fe/H] = 0.06. Since no elemental abundances beyond [Fe I \& II/H] have been determined for W23375 ([Fe/H] = -0.18), we draw no conclusions for any relationship that may exist between that star and W12388. Turning to the three stars in the [Fe/H]$\sim$ -0.50 range, with the exception of [Ni/Fe] Figure \ref{f4} shows $\sim$0.25 $\leq$ [X/Fe] $\leq$ 0.61 dex dispersion in measured abundance ratios for all other elements considered in this study. This large element--by--element abundance scatter is incompatible with these stars being chemically related.\\

\subsection{W11449 \& W11450}

The two metal--rich stars in this study, W11449 \& W11450, were first noted by \citet{1932ngcd.book.....A} in a catalog of double stars located about the North Galactic pole. LMS91 lists an 8.0$^{\prime\prime}$ angular separation on the sky for the stars, with a physical separation of $\sim$600 AU and $P_{bound}$ = 0.85. The detailed kinematic study proposes these stars are wide binary candidates. Although beyond the scope of this study, \citet{2010MNRAS.404.1835K} suggests that wide binaries may be formed during the dissolution of a star cluster, leaving open the prospect of finding isolated, chemically related wide binaries in the Galactic field. \\

\begin{figure}
\caption{A comparison of measured abundances for W11449 \& W11450. Both stars exhibit similar abundance patterns, element by element, suggesting these stars may be chemically related.}
\epsscale{1.00}
\hspace{-0.1 in}
\plotone{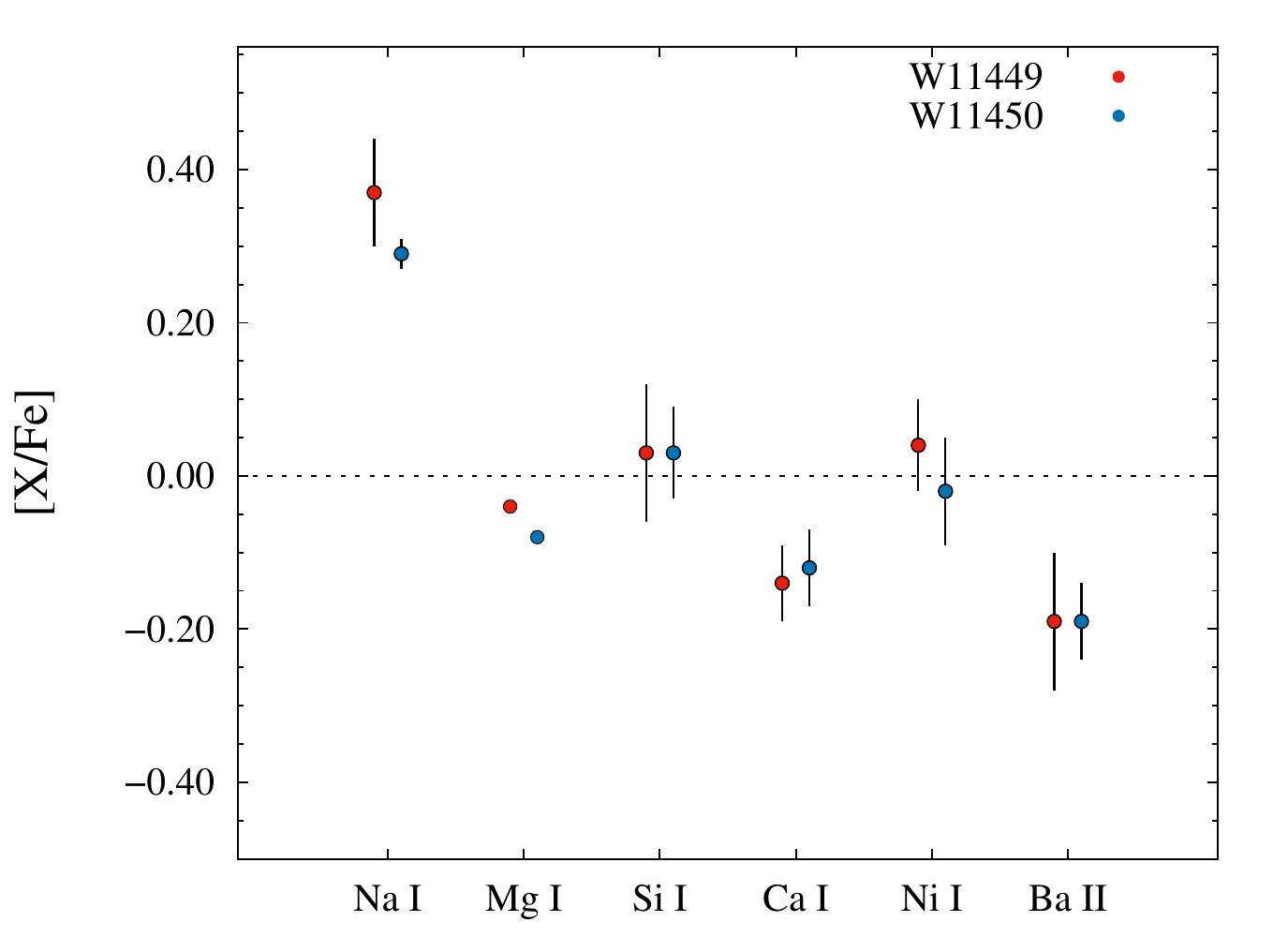}
\label{f5}
\end{figure}

\begin{figure}
\caption{Measured abundances relative to iron for six stars (W23375 excluded) in Latham 1. [Na/Fe] is enhanced in all program stars, except for the more metal--poor W23508, but especially for the two metal--rich stars, W11449 \& W11450.}
\epsscale{1.10}
\hspace{0.65in}
\plotone{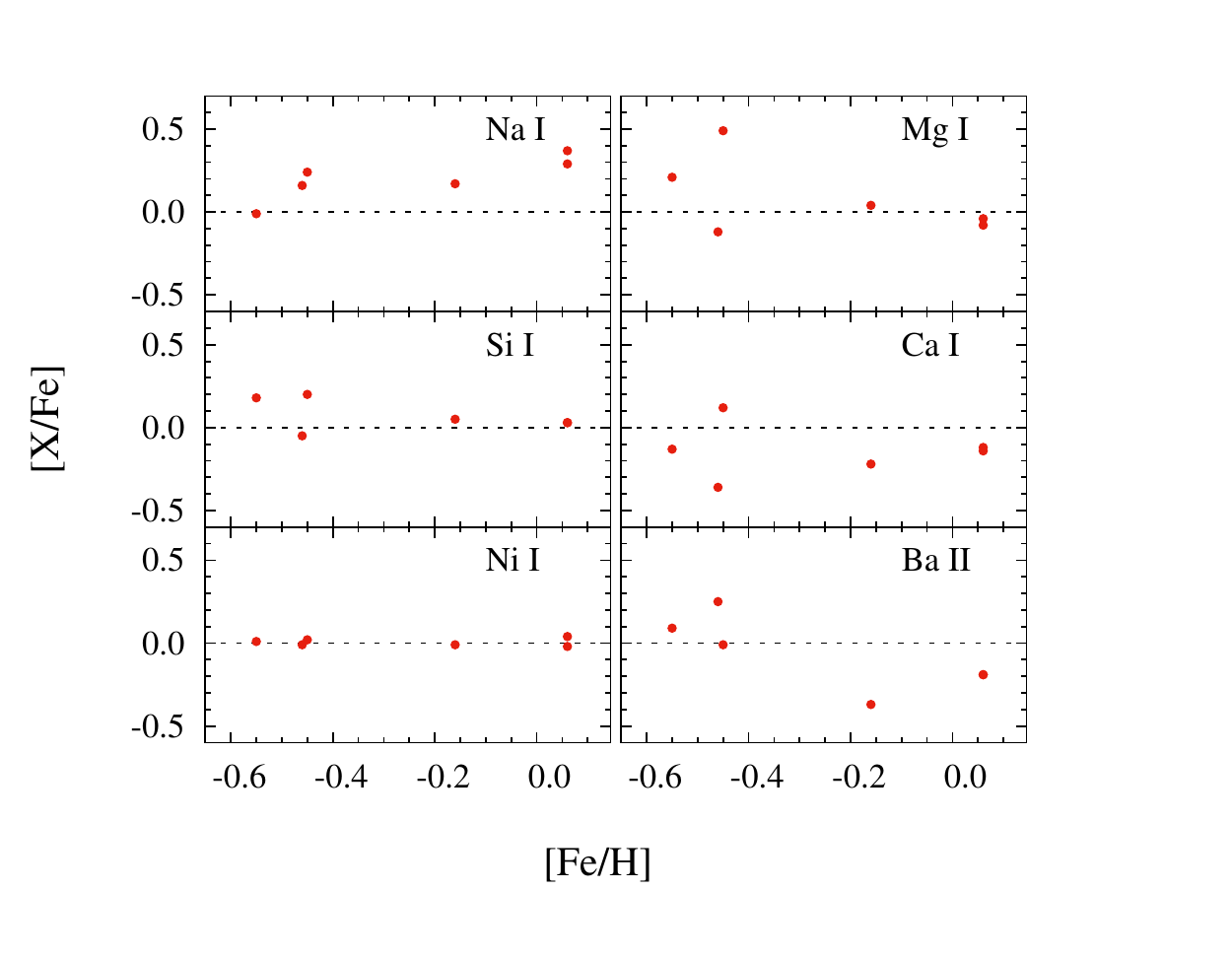}
\label{f6}
\end{figure}

Beyond the near solar metallicity for these stars, [Fe/H] = 0.06 for both W11449 \& W11450, we find a notable similarity in abundance ratios for all other elements considered in this study and displayed in Figure \ref{f5}. We see the greatest difference between the two stars in Na I. [Na/Fe] is noticeably enhanced by $\langle$0.33$\rangle$ dex for both stars with $\Delta_{\rm [Na/Fe]}$ = 0.08. Both \citet{1993A&A...275..101E} and \citet{2000A&A...363..692T} cite an increase in Na I with [Fe/H], which can be seen in the ``upturn" in the first panel of Figure \ref{f6}. A compilation of Galactic cluster abundances  by \citet{2006cams.book....3F} reveals this is not a peculiar over--abundance for stars in the solar neighborhood. The $\alpha$-- elements, Mg, Si and Ca appear to follow the same abundance trends as disk field stars as well  \citep[see also][for field star trends in Ca at this metallicity]{2004MmSAI..75..267S}. We conclude these two stars show a striking similarity in abundances for all elements presented in this study, and may very well have shared the same stellar formation history.\\

\section{SUMMARY}

We present the first high resolution (R $\sim$ 60,000) high SNR ($\sim\langle$100$\rangle$) spectroscopic abundance analysis for all seven stars in Moving Group W11450 (Latham 1) to determine if they may be chemically related. For six stars we have derived  Fe I \& Fe II (neutral and singly ionized iron, respectively), Na I, Mg I, Si I, Ca I, Ni I and Ba II. For the warmest star in the group, W23375 ($\sim$7000 K), we present results for Fe abundances only. All data for this study was obtained at McDonald Observatory with the Otto Struve 2.1m telescope and Sandiford Cass Echelle Spectrograph (SES) in June of 2013 and 2014. The spectrograph setup was centered near 6175~\AA\ in 2013 and  6375~\AA\ in 2014 with wavelength coverage spanning from $\sim$5350--6560~\AA\ and $\sim$5560--6750~\AA, respectively.  Program stars are located near the North Galactic Cap in Selected Area 57 (SA 57; \emph{l}=66$\degr$, \emph{b}=+86$\degr$) and range in magnitude from $8.25 \leq m_V\leq 11.47$ after correcting to Johnson photometric magnitudes from Tycho-- 2 colors.\\

Effective temperatures (T$_{\rm eff}$) for individual stars were initially estimated by utilizing the Tycho--2 $(B_T-V_T)$ color indices converted to the Johnson photometric system. The correction for interstellar reddening and extinction was performed to render the appropriate $(B-V)_0$ color index for each star. Surface gravities (log $g$) were initially estimated by applying a $V$--magnitude bolometric correction to the absolute $V$--band magnitudes estimated from the distance moduli for individual stars.  An iterative LTE stellar line analysis code (MOOG) was used to further refine model atmosphere parameters, T$_{\rm eff}$, log $g$, and microturbulence ($\xi$), and to derive elemental abundances via equivalent width (EW) analysis. Final T$_{\rm eff}$ were modeled by removing trends in Fe I with excitation potential. Surface gravities were further constrained by Fe ionization equilibrium. We initially assumed a $\xi$ value of 2 km $\rm s^{-1}$ for all stars, and achieved a final value by eliminating any remaining bias in Fe I and reduced EW [log(EW/$\lambda$)]. Elemental abundances were determined when a final solution over all parameters was achieved.\\

Even though Latham 1 is listed in the Dias catalog of open clusters, our analysis finds a large  star--to--star scatter in metallicity, -0.55 $\leq$ [Fe/H] $\leq$ 0.06 ($\sigma$= 0.25), inconsistent with all stars being formed from the same chemically homogeneous molecular cloud. An examination of the $>\,\sim$0.60 dex spread in [Fe/H] ratios appears to separate the stars into three distinct groups:  [Fe/H]$\sim$ -0.17, [Fe/H]$\sim$ -0.50, and [Fe/H] = 0.06. Since no elemental abundances beyond [FeI \& II/H] were determined for W23375 ([Fe/H] = -0.18), we draw no conclusions for any relationship that may exist between that star and W12388. Three stars in the [Fe/H]$\sim$ -0.50 range, with the exception of [Ni/Fe], show $\sim$0.25 $\leq$ [X/Fe] $\leq$ 0.61 dex dispersion in measured abundance ratios for all other elements considered in this study. This large element--by--element abundance scatter is incompatible with these stars being chemically related.\\

However, for the two remaining stars in this study, wide binary candidates W11449 \& W11450, we find almost solar [Fe/H] = 0.06 ratios and a small spread in abundance ratios with respect to iron for all other elements considered in this study. Both stars are noticeably Na enhanced by $\langle$0.33$\rangle$ dex and this element shows the largest spread with $\Delta_{\rm [Na/Fe]}$ = 0.08 for the two stars. The $\alpha$-- elements, Mg, Si and Ca, appear to follow Galactic abundance trends as other disk field stars, with $\langle$[$\alpha$/Fe]$\rangle$ = -0.04. Ba II abundances appear to be deficient with respect to solar barium abundance. Our analysis suggests these two stars may share a common stellar formation history, and that the previously proposed connection between these two stars may be real.\\

\section*{Acknowledgments}

We thank the referees for helpful suggestions that have improved this paper immensely. We extend gratitude to Matthew Shetrone for his help with observations, and to both he and Chris Sneden for helpful discussions with the analysis. We would also thank John Donor for his help confirming distance estimates for these stars. This research has made use of the NASA/IPAC Extragalactic Database (NED) which is operated by the Jet Propulsion Laboratory, California Institute of Technology, under contract with the National Aeronautics and Space Administration. This work was supported in part by the National Science Foundation through an REU Site Program grant to Texas Christian University, PHY--1358770 for KM. Support by the College of Science and Engineering at Texas Christian University for JEO is gratefully acknowledged. This material is based upon work supported by the National Science Foundation under award AST--1311835 to PMF.
\vspace{5mm}
\facilities{Struve (Sandiford Echelle Spectrograph)}

\software{IRAF}

\bibliography{Bibliography}

\clearpage
\begin{deluxetable*}{lcccccccccc}
\tablecolumns{11}
\tablewidth{0pt}
\tabletypesize{\scriptsize}
\tablecaption{\label{tab1}Summary from Previous Studies}
\tablehead{
           \multicolumn{2}{c}{}               &
           \multicolumn{2}{c}{Latham et al.}             &
           \multicolumn{4}{c}{Ammons et al.}              &
           \multicolumn{1}{c}{Casagrande et al.}               &
           \multicolumn{2}{c}{Pickles \& Depagne}       \\ 
\cmidrule(lr){3-4} \cmidrule(lr){5-8} \cmidrule(lr){9-9} \cmidrule(lr){10-11}
\noalign{\smallskip}
\colhead{Star ID\,\tablenotemark{a}}	&
\colhead{Alt ID\,\tablenotemark{b}}	&
\colhead{distance\,\tablenotemark{c}}      &
\colhead{Spec. Type\,\tablenotemark{d}}      &
\colhead{T$\rm_{eff}$\,\tablenotemark{e}}      &
\colhead{[Fe/H]\,\tablenotemark{f}}      &
\colhead{$\sigma$}      &
\colhead{distance\,\tablenotemark{g}}      &
\colhead{[Fe/H]\,\tablenotemark{h}}        &
\colhead{distance\,\tablenotemark{i}}      &
\colhead{Spec. Type\,\tablenotemark{j}}    \\
\colhead{(Weistrop)}	&
\colhead{(Tycho--2)}	&
\colhead{(pc)}      &
\colhead{}             &
\colhead{(K)}	&
\colhead{}      &
\colhead{}      &
\colhead{(pc)}      &
\colhead{}      &
\colhead{(pc)}      &
\colhead{}           
}

\startdata				  
W11449 & 2532-2234-1 & 82 & G7 V & 5453 & \ldots &\ldots &     66  & \ldots & 143 & G8 V  \\
W11450 & 2532-2235-1 & 76 & G4 V & 5696 & 0.70 & $\pm{0.20}$ &     47  & \ldots   & 128 & rG5 V   \\
W12388 & 2532-1132-1 & \ldots & \ldots & 5926 & 0.90 & $\pm{2.10}$  & 29  & \ldots & 306 & rG0 V  \\
W13284 & 2535-0716-1 & \ldots &\ldots & 8146  & \ldots & \ldots & 66  & \ldots & 418 & F0 V  \\
W21415 & 2532-0862-1 & \ldots & \ldots & 5350 & \ldots & \ldots & 108 & \ldots & 294 & rG5 V  \\
W23375 & 2532-1161-1 & 101  & F4 V & 6520 & 0.18 & $\pm^{0.12}_{0.13}$ & 87  & $-$0.06  & 77 & wF5 V   \\
W23508 & 2532-1163-1 & 164 & G1 V & 5992 & 0.00 & $\pm^{0.33}_{0.34}$ & 47 & \ldots & 231 & rF8 V   \\
\enddata
\tablenotetext{a}{\,\,\,\,\,\,\,\,\,\,\,\,\,\,Star Identifiers from the Weistrop catalog (1980) used by Lathem et al.}
\tablenotetext{b}{\,\,\,\,\,\,\,\,\,\,\,\,\,\,Alternate star identifiers found in the Tycho--2 catalog.}
\tablenotetext{c,d}{\,\,\,\,\,\,\,\,\,\,\,\,\,Distances and spectral types from \citet{1991AJ....101..625L}.}
\tablenotetext{e,f,g}{\,\,\,\,\,\,\,\,\,\,\,\,\,Photometric estimates of T$_{\rm eff}$ and [Fe/H] with uncertainties, and distances from \citet{2006ApJ...638.1004A}.}
\tablenotetext{h}{\,\,\,\,\,\,\,\,\,\,\,\,\,Photometric estimate of [Fe/H] from \citet{Casagrande}.}
\tablenotetext{i,j}{\,\,\,\,\,\,\,\,\,\,\,\,\,Distances and spectral types from \citet{2010PASP..122.1437P}. An ``r" or ``w" preceding the spectral type\\ indicates a ``rich" or ``weak" metallicity with respect to solar.}

\end{deluxetable*}


\begin{deluxetable*}{lccccrc}
\tablecolumns{7}
\tablewidth{0pt}

\tabletypesize{\normalsize}
\tablecaption{\label{tab2}Observation Logs, 2013 and 2014}
\tablehead{
\colhead{Star ID\,\tablenotemark{a}} &
\colhead{Alt ID\,\tablenotemark{b}} &
\colhead{Date of Observations}   &
\colhead{RA (J2000)\,\tablenotemark{c}} &
\colhead{Dec}      &
\colhead{$m_V$} &
\colhead{ Exposure Time}      \\
\colhead{(Weistrop)} &
\colhead{(Tycho--2)} &
\colhead{}   &
\colhead{(hr:min:sec)}   &
\colhead{(deg:min:sec)}      &
\colhead{}      &
\colhead{(sec)}  
}
  
\startdata
W11449 & 2532-2234-1 & 2014-06-11,13      & 13:11:12.025 & +30:49:37.85 & 10.99  & 4$\times$1200\\
W11450 & 2532-2235-1 & 2014-06-12,15      & 13:11:12.681 & +30:49:36.93 & 9.54 & 3$\times$1200\\
W12388 & 2532-1132-1 & 2014-06-11,15,17  & 13:11:36.441 & +30:19:45.12 & 11.26 & 4$\times$1200\\
W13284 & 2535-0716-1 & 2014-06-10,13      & 13:12:54.003 & +30:01:03.42 & 10.98  & 4$\times$1200\\
W21415 & 2532-0862-1 & 2013-06-22,23      & 13:09:22.779 & +30:54:15.94 & 11.47  & 4$\times$1200\\
W23375 & 2532-1161-1 & 2013-06-24           & 13:09:06.493 & +30:14:37.46 & 8.25  & 2$\times$900\\
W23508 & 2532-1163-1 & 2013-06-24           & 13:10:21.598 & +30:11:24.58 & 10.63  & 4$\times$1200\\
\enddata
\tablecomments{Tycho-- 2 photometry is converted to the Johnson photometric system (see \S 2 for specifics).}
\tablenotetext{a,b}{\,\,\,\,\,\,\,\,Star identifiers are the same as Table \ref{tab1}.}
\tablenotetext{c}{\,\,\,\,\,\,\,\,Updated J2000 coordinates from SIMBAD astronomical database.}
\end{deluxetable*}


\begin{deluxetable*}{ccccccc}
\tablecolumns{7}
\tablewidth{0pt}
\tablecaption{\label{tab3}Measured Radial Velocities}

\tablehead{
  \colhead{Star\,\tablenotemark{a}} &
  \colhead{$V_{\rm rad}$} &
  \colhead{$\sigma_{V_{rad}}$} &
  \colhead{Latham\,\tablenotemark{b}} &
  \colhead{$\sigma_{V_{rad}}$} &
  \colhead{Stefanik\,\tablenotemark{c}} &
  \colhead{$\sigma_{V_{rad}}$}  \\ 
  \colhead{} &
  \colhead{(km $\rm s^{-1}$)} &
  \colhead{} &
  \colhead{ (km $\rm s^{-1}$)} &
  \colhead{} &
  \colhead{ (km $\rm s^{-1}$)} &
  \colhead{}  
}
\startdata
  W11449 & $-$2.16 & 0.48 & $-$2.12 & 0.08 & \nodata & \nodata\\
  W11450 & $-$2.56 & 0.54 & $-$2.40 & 0.50 & \nodata & \nodata\\
  W12388 & $-$3.06 & 0.54 & $-$2.56 & 0.16 & $-$3.07 & 0.16\\
  W13284 & $-$2.53 & 0.87 &  $-$2.51 & 0.21 & $-$2.48 & 0.16\\
  W21415 & $-$2.56 & 0.45 & $-$2.90 & 0.28 & $-$2.99 & 0.20\\
  W23375 & $-$2.07 & 1.22 & $-$2.12 & 0.25 & \nodata & \nodata\\
  W23508 & $-$2.68 & 0.62 & $-$2.30 & 0.24 & $-$2.75 & 0.18\\
\enddata
\tablecomments{For W11450 Lathem et al. estimates the center of mass velocity for a possible tertiary system.}
\tablenotetext{a}{\,Star Identifiers from the Weistrop catalog (1980). Refer to Table \ref{tab1} for Tycho--2 catalog star ID's.}
\tablenotetext{b}{\,Radial velocities from \citet{1991AJ....101..625L}.}
\tablenotetext{c}{\,Radial velocities from \citet{2006PASP..118.1656S}.}

\end{deluxetable*}

\begin{deluxetable*}{lrccrrrr}
\tablecolumns{8}
\tablewidth{0pt}

\tablecaption{\label{tab4}Final Model Atmospheres and Photometry}
\tablehead{
\colhead{Star\,\tablenotemark{a}}	&
\colhead{T$\rm_{eff}$}      &
\colhead{log $g$}      &
\colhead{$\xi$}      &
\colhead{[Fe/H]}      &
\colhead{$m_V$}      &
\colhead{B-V}      &
\colhead{SNR}        \\
\colhead{}	&
\colhead{(K)}	&
\colhead{(cgs)}      &
\colhead{(km s$\rm ^{-1}$)}      &
\colhead{}      &
\colhead{}      &
\colhead{}      &
\colhead{}
}

\startdata				  
W11449 & 5575 & 4.4 & 1.8 & 0.06  & 10.99 & 0.66 & 90  \\
W11450 & 5700 & 4.2 & 1.6 & 0.06  & 9.54 & 0.67 & 120  \\
W12388 & 5650 & 4.0 & 1.8 & $-$0.16  & 11.26 & 0.75 & 100  \\
W13284 & 6450 & 3.5 & 1.8 & $-$0.46  & 10.98 & 0.40 & 110  \\
W21415 & 5770 & 4.1 & 1.4 & $-$0.45 & 11.47 & 0.57 & 100  \\
W23375 & 6990 & 4.5 & 2.0 & $-$0.18  & 8.25 & 0.41 & 70  \\
W23508 & 6300 & 4.4 & 1.8 & $-$0.55 & 10.63 & 0.54 & 130  \\
\enddata
\tablecomments{Tycho-- 2 photometry is converted to the Johnson photometric system (see \S 2 for specifics).}
\tablenotetext{a}{\,Star Identifiers from the Weistrop catalog (1980). Refer to Table \ref{tab1} for Tycho--2 catalog star ID's.}
\end{deluxetable*}

\clearpage
\begin{deluxetable*}{ccccccccccc}
\tablecolumns{11}
\tablewidth{0pt}
\tablecaption{\label{tab5}Atomic Data and Equivalent Widths}
\tablehead{
\colhead{Wavelength}	&
\colhead{Species}      &
\colhead{E.P.}      &
\colhead{log $gf$}      &
\colhead{W11449}      &
\colhead{W11450}      &
\colhead{W12388}      &
\colhead{W13284}      &
\colhead{W21415}      &
\colhead{W23375}      &
\colhead{W23508}      \\
\colhead{(\AA)}	&
\colhead{}	&
\colhead{(eV)}      &
\colhead{}      &
\colhead{(m\AA)}      &
\colhead{(m\AA)}      &
\colhead{(m\AA)}      &
\colhead{(m\AA)}      &
\colhead{(m\AA)}      &
\colhead{(m\AA)}      &
\colhead{(m\AA)}
}

\startdata
  5441.339  &   Fe I  &  4.312  &  $-$1.73  &  \nodata &  \nodata &  \nodata &  \nodata &  13.10 & \nodata & 6.30 \\
  5445.042  &   Fe I  &  4.386  &  $-$ 0.21 &  \nodata & \nodata & \nodata & \nodata & 85.90  & 74.50 & 68.30 \\
  5576.089  &   Fe I  &  3.430  &  $-$0.94  & 161.10 & 136.60 & 128.00 &74.70 & 91.70 &  73.80 & \nodata \\   
  5638.262  &   Fe I  &  4.220  &  $-$0.72  & \nodata & \nodata &  89.80 &  \nodata &  56.20 & \nodata &  \nodata \\  
  5679.024  &   Fe I  &  4.652  &  $-$0.90  &  72.20 &  63.90 &  63.00 & 23.70 &  36.10 &  21.60 & 23.10 \\  
  5752.032  &   Fe I  &  4.548  &  $-$0.86  &  79.20 & 74.60 &  \nodata &  \nodata &  38.30 & \nodata &  20.30 \\  
  5760.344  &   Fe I  &  3.642  &  $-$2.44  &  40.50 &  29.90 & 24.60 &  \nodata &  12.40 & \nodata & \nodata\\  
  5775.081  &   Fe I  &  4.220  &  $-$1.30  &  78.30  &  69.30 & 59.10 &  \nodata &  35.90 & \nodata &  16.50\\  
  5793.915  &   Fe I  &  4.220  &  $-$1.66  &  54.60 &  41.30 &  40.40 &  \nodata &  16.50 & \nodata & 10.40 \\  
  5806.726  &   Fe I  &  4.607  &  $-$1.03  &  67.10 &  58.70 &  49.00 & 17.40 &  33.10 &  18.40 &  16.90 \\ 
  5855.076  &   Fe I  &  4.604  &  $-$1.48  &  35.00 &  \nodata &  21.70 &  \nodata & \nodata & \nodata & \nodata\\  
  5856.088  &   Fe I  &  4.294  &  $-$1.64  &  50.30 &  44.30 &  30.50 &  \nodata & 14.30 & \nodata & \nodata\\  
  5916.247  &   Fe I  &  2.453  &  $-$2.99  &  81.80 &  68.70 &  59.80 &  \nodata &  39.40 & \nodata & 19.00 \\  
  5927.789  &   Fe I  &  4.652  &  $-$1.07  &  71.10 &  60.80 &  \nodata &  \nodata &  \nodata & \nodata & \nodata\\  
  5929.677  &   Fe I  &  4.548  &  $-$1.41  &  \nodata & \nodata &  \nodata &  \nodata &  17.70 & \nodata & \nodata\\  
  5934.655  &   Fe I  &  3.928  &  $-$1.12  & 102.20 &  96.10 & \nodata & 35.20 &  53.50 &  38.50 &  32.60 \\  
  5987.065  &   Fe I  &  4.795  &  $-$0.43  &  94.30 &  87.40 & \nodata & 32.10 &  \nodata & \nodata & \nodata\\ 
  6056.005  &   Fe I  &  4.733  &  $-$0.46  &  97.60 &  89.70 &  76.00 & 33.40 &  51.10 &  39.90 &  34.60 \\  
  6079.008  &   Fe I  &  4.652  &  $-$1.10  &  68.40 &  \nodata &  43.20 &  \nodata &  25.90 & \nodata &  12.50 \\  
  6165.360  &   Fe I  &  4.143  &  $-$1.55  &  57.60 &  58.10 &  45.30 &  \nodata &  28.00 & \nodata &  15.30 \\  
  6180.203  &   Fe I  &  2.727  &  $-$2.65  &  86.30 &  75.70 & 61.60 &  \nodata &  37.60 & \nodata &  15.60 \\  
  6187.989  &   Fe I  &  3.943  &  $-$1.67  &  71.50 &  61.50 &  46.60 &  \nodata &  26.60 & \nodata &  12.50 \\ 
  6200.313  &   Fe I  &  2.608  &  $-$2.44  & 111.00 &  92.10 &  89.50 &  \nodata &  56.10 &  26.60 &  30.20 \\  
  6229.226  &   Fe I  &  2.845  &  $-$2.81  &  63.10 &  56.90 & 43.50 &  \nodata &  22.00 & \nodata & \nodata\\  
  6232.640  &   Fe I  &  3.654  &  $-$1.24  & 115.60 & 101.60 &  \nodata & 42.10 &  \nodata & \nodata &  41.90 \\ 
  6246.320  &   Fe I  &  3.602  &  $-$0.77  & 167.90 & 142.30 & 126.20 & 71.70 & 101.10 & \nodata &  80.10 \\   
  6270.223  &   Fe I  &  2.858  &  $-$2.71  &  69.00 &  57.50 &  55.90 & 12.70 &  30.40 & \nodata &  17.30 \\ 
  6301.501  &   Fe I  &  3.654  &  $-$0.71  & \nodata & 147.60 & 125.00 &  73.90 & \nodata & \nodata & 88.50 \\  
  6322.685  &   Fe I  &  2.588  &  $-$2.43  &  \nodata &  \nodata &  \nodata &  \nodata &  57.80 & \nodata &  30.80 \\  
  6336.824  &   Fe I  &  3.686  &  $-$0.85  &  153.30 &  \nodata &  \nodata & 63.00 &  85.70 & \nodata &  59.00 \\  
  6393.600  &   Fe I  &  2.433  &  $-$1.62  &  \nodata &  \nodata &  \nodata & 85.20 &  \nodata  &  84.20 &  82.70 \\  
  6592.913  &   Fe I  &  2.727  &  $-$1.47  & \nodata & 156.80 & 145.10 & 81.30 &  \nodata & \nodata & \nodata\\  
  6593.870  &   Fe I  &  2.433  &  $-$2.42  & 121.40 & 105.30 & 102.30 & 39.80 &  \nodata & \nodata & \nodata\\  
  5425.257  &   Fe II  &  3.199  &  $-$3.35  &  \nodata  &  \nodata &  \nodata &  \nodata&  28.00 & \nodata &  24.30\\  
  6084.102  &   Fe II  &  3.199  &  $-$3.85  &  22.00 &  \nodata &  22.50 & 23.90 &  \nodata & \nodata & \nodata\\  
  6149.246  &   Fe II  &  3.889  &  $-$2.73  &  \nodata &  45.70 &  42.30 & 50.80 &  \nodata & \nodata &  26.40\\  
  6247.557  &   Fe II  &  3.891  &  $-$2.34  &  \nodata & \nodata &  63.50 &  \nodata &  42.30 &  67.30 & \nodata\\  
  6369.462  &   Fe II  &  2.891  &  $-$4.25  &  19.10 &  \nodata &  \nodata &  \nodata &  \nodata & \nodata & \nodata\\  
  6416.919  &   Fe II  &  3.891  &  $-$2.74  &  45.50 &  44.20 &  45.00 & 47.30 &  28.90 &  46.90 &  26.60\\  
  6432.679  &   Fe II  &  2.891  &  $-$3.75  &  38.70 &  46.60 &  \nodata & 46.40 &  27.90 & \nodata & \nodata\\  
  6456.381  &   Fe II  &  3.903  &  $-$2.09  &  \nodata &  86.40 &  \nodata &  \nodata &  \nodata & \nodata &  59.10\\  
  6154.226  &   Na I  &  2.102  &  $-$1.55  &  90.10 &  63.10 &  39.80 & 11.70 &  26.70 & \nodata &  8.90 \\  
  6160.747  &   Na I  &  2.104  &  $-$1.25  & 107.80 &  84.90 &  63.40 & 23.10 &  39.90 & \nodata &  14.80 \\  
  5711.088  &   Mg I  &  4.346  &  $-$1.72  & 142.00 & 121.60 & 111.20 & 51.00 & 115.20 & \nodata &  70.50\\  
  6318.717  &   Mg I  &  5.108  &  $-$2.10  & \nodata &  \nodata &  38.10 &  \nodata &  43.70 & \nodata &  17.20\\  
  6319.237  &   Mg I  &  5.108  &  $-$2.32  &  \nodata &  \nodata &  \nodata &  \nodata &  \nodata & \nodata & \nodata\\  
  6319.495  &   Mg I  &  5.108  &  $-$2.80  &  \nodata &  \nodata &  \nodata &  \nodata & \nodata & \nodata & \nodata\\  
  5690.425  &   Si I  &  4.929  &  $-$1.87  &  \nodata &  \nodata & \nodata & \nodata &  34.10 & \nodata & \nodata\\  
  5701.104  &   Si I  &  4.929  &  $-$2.05  &  53.30 &  48.90 &  \nodata &  \nodata & \nodata & \nodata &  17.10 \\  
  5772.146  &   Si I  &  5.082  &  $-$1.36  &  \nodata & \nodata & \nodata &  34.40 &  \nodata & \nodata &  \nodata \\  
  5793.073  &   Si I  &  4.929  &  $-$2.06  &  \nodata & \nodata & \nodata &  \nodata &  \nodata & \nodata & \nodata \\  
  6142.483  &   Si I  &  5.619  &  $-$1.30  &  49.60 & 55.70 &  38.20 &  \nodata &  \nodata & \nodata & \nodata \\  
  6145.016  &   Si I  &  5.616  &  $-$1.31  &  53.00 &  50.30 &  45.10 & \nodata & \nodata & \nodata & \nodata\\  
  6155.134  &   Si I  &  5.619  &  $-$0.76  & 121.90 & 110.30 &  91.30 & 36.40 & 72.20 & \nodata & 49.90 \\ 
  6243.815  &   Si I  &  5.616  &  $-$1.24  &  \nodata & \nodata & \nodata & \nodata & 36.40 & \nodata & \nodata\\  
  5581.965  &   Ca I  &  2.523  &  $-$0.71  & 121.50 & 99.80 & 85.40 &  \nodata & 81.50 & \nodata & \nodata\\  
  6169.042  &   Ca I  &  2.523  &  $-$0.55  & 138.00 & 121.00 &  96.10 & 36.00 & 85.90 & \nodata &  54.90 \\  
  6169.563  &   Ca I  &  2.526  &  $-$0.27  & \nodata &  \nodata & 112.50 & 59.70 &104.80 & \nodata &  72.50 \\  
  6471.662  &   Ca I  &  2.526  &  $-$0.59  & 126.90 & 117.50 & 105.70 & \nodata &  93.50 & \nodata &  \nodata \\   
  5593.732  &   Ni I  &  3.898  &  $-$0.84  &  61.80 &  50.60 &  45.60 &  \nodata &  24.50 & \nodata &  13.60 \\ 
  5760.828  &   Ni I  &  4.105  &  $-$0.80  &  \nodata &  \nodata &  33.20 &  \nodata &  21.60 & \nodata & \nodata\\  
  5805.213  &   Ni I  &  4.167  &  $-$0.58  &  59.30 & 53.10 &  38.50 & 12.10 &  23.00 & \nodata &  12.10 \\  
  6176.812  &   Ni I  &  4.088  &  $-$0.26  &  84.70 &  72.40 &  65.20 & 28.10 &  41.90 & \nodata &  26.40 \\  
  6204.604  &   Ni I  &  4.088  &  $-$1.08  &  39.70 &  33.20 &  21.70 &  \nodata &  \nodata & \nodata & \nodata\\  
  6223.984  &   Ni I  &  4.105  &  $-$0.86  &  45.80 &  34.70 &  29.40 &  \nodata &  \nodata & \nodata & \nodata\\   
  5853.668  &   Ba II  &  0.604  &  $-$1.01  &  70.90 &  73.90 & 60.10 & 72.70 & 51.80 & \nodata &  47.40 \\  
  6141.713  &   Ba II  &  0.704  &  $-$0.07  & 131.90 & 115.40 & 107.20 & 122.70 & 94.50 & \nodata &  95.20 \\  
  6496.897  &   Ba II  &  0.604  &  $-$0.38  & \nodata &  \nodata & \nodata & \nodata & 87.70 & \nodata & \nodata\\  
\enddata


\end{deluxetable*}

\newpage
\floattable
\begin{deluxetable*}{lccccccc}
\tabletypesize{\small}
\rotate
\tablecolumns{8} 
\tablewidth{0pt} 
\tablecaption{\label{tab6}Measured Abundances}
\tablehead{\colhead{Species}                                    &
           \colhead{W11449}                             &
           \colhead{W11450}                   	      &
           \colhead{W12388}                             &
           \colhead{W13284}                             &
           \colhead{W21415}                   	      &
           \colhead{W23375}                             &
           \colhead{W23508}                             \\       
           \colhead{}                         		      &
           \colhead{} 				      &
           \colhead{}                                          &               
           \colhead{}                                          &            
           \colhead{}                                          & 
           \colhead{}                                          & 
           \colhead{}                                          &
           \colhead{}                                          }
\startdata
\lbrack{FeI/H}\rbrack     & \phantom{$-$}0.06  $\pm$ 0.06 (25)& \phantom{$-$}0.06 $\pm$ 0.04 (24)&  $-$0.16 $\pm$ 0.06 (22)    &  $-$0.46 $\pm$ 0.05 (14)   &  $-$0.45 $\pm$ 0.06 (25) &  $-$0.18 $\pm$ 0.02 (8)  &  $-$0.55 $\pm$ 0.08 (22) \\
\lbrack{FeII/H}\rbrack   &  \phantom{$-$}0.06  $\pm$ 0.05 (4) & \phantom{$-$}0.04 $\pm$ 0.07 (4)    &  $-$0.16 $\pm$ 0.03 (4)  &  $-$0.45 $\pm$ 0.04 (4)  &  $-$0.45 $\pm$ 0.06 (4)  &  $-$0.19 $\pm$ 0.04 (2) &  $-$0.55 $\pm$ 0.01 (4)\\
\lbrack{Na/Fe}\rbrack   & \phantom{$-$}0.37 $\pm$ 0.07 (2)& \phantom{$-$}0.29$\pm$ 0.01 (2)& \phantom{$-$}0.17 $\pm$ 0.03 (2) & \phantom{$-$}0.16 $\pm$ 0.03 (2)& \phantom{$-$}0.24 $\pm$ 0.04 (2) & \ldots & $-$0.01 $\pm$ 0.03 (2) \\ 
\lbrack{Mg/Fe}\rbrack   & $-$0.04 $\pm$ 0.00 (1)               & $-$0.08 $\pm$ 0.00  (1)& \phantom{$-$}0.04 $\pm$ 0.07 (2)&  $-$0.12 $\pm$ 0.00 (1) & \phantom{$-$}0.49 $\pm$ 0.05  (2)& \ldots  &   \phantom{$-$}0.21 $\pm$ 0.08 (2)    \\
\lbrack{Si/Fe}\rbrack     & \phantom{$-$}0.03 $\pm$ 0.09 (4)&\phantom{$-$}0.03 $\pm$ 0.06 (4)  &  \phantom{$-$}0.05 $\pm$ 0.07 (3) &  $-$0.05 $\pm$ 0.08 (2) & \phantom{$-$}0.20 $\pm$ 0.02 (3) & \ldots &  \phantom{$-$}0.18 $\pm$ 0.08 (2) \\
\lbrack{Ca/Fe}\rbrack   & $-$0.14 $\pm$ 0.05  (3) &  $-$0.12$\pm$ 0.05  (3)                             &  $-$0.22 $\pm$ 0.09 (4)  &  $-$0.36 $\pm$ 0.08 (2)  &  \phantom{$-$}0.12 $\pm$ 0.07 (4) & \ldots  &  $-$0.13 $\pm$ 0.01 (2) \\
\lbrack{Ni/Fe}\rbrack    &  \phantom{$-$}0.04 $\pm$ 0.06  (5) &  $-$0.02 $\pm$ 0.07  (5)& $-$0.01 $\pm$ 0.04 (6)   &  $-$0.01 $\pm$ 0.04 (2)  &   \phantom{$-$}0.02 $\pm$ 0.05  (4)     & \ldots         &   \phantom{$-$}0.01 $\pm$ 0.05 (3)    \\
\lbrack{BaII/Fe}\rbrack & $-$0.19 $\pm$ 0.09  (2)               & $-$0.19 $\pm$ 0.05  (2)& $-$0.37 $\pm$ 0.01 (2)    & \phantom{$-$}0.25 $\pm$ 0.09 (2)    & $-$0.01 $\pm$ 0.05 (3)  & \ldots  &   \phantom{$-$}0.09 $\pm$ 0.01  (2)   \\
\enddata 
\vspace{0.05cm}
\tablecomments{Number of absorption lines used in each abundance analysis is indicated in parentheses.}
\end{deluxetable*}

\clearpage

\newpage

\newpage

\end{document}